\documentstyle[aps,prl,floats,epsfig]{revtex}

\title{Crossover from Electronic to Atomic Shell Structure in Alkali Metal Nanowires}
\author{A.I. Yanson$^{1}$\thanks{Present address: Dept. of Physics,
510 Clark Hall, Cornell University, Ithaca, NY 14853} 
, I.K. Yanson$^{1, 2}$, and J.M. van Ruitenbeek$^{1}$}
\address{$^{1}$Kamerlingh Onnes Laboratorium, Universiteit Leiden, \\
Postbus 9504, NL-2300 RA Leiden, The Netherlands\\
$^{2}$B. Verkin Institute for Low Temperature Physics and Engineering,\\
National Academy of Sciences, 310164, Kharkiv, Ukraine}

\begin{document}
\draft

\twocolumn[\hsize\textwidth\columnwidth\hsize\csname@twocolumnfalse\endcsname

\maketitle

\begin{abstract}
{After making a cold weld by pressing two clean metal surfaces together, upon gradually separating the two pieces a metallic nanowire is formed, which progressively thins down to a single atom before contact is lost.  In previous experiments \cite{yyr1,yyr2} we have observed that the stability of such nanowires is influenced by electronic shell filling effects, in analogy to shell effects in metal clusters \cite{deheer93}. For sodium and potassium at larger diameters there is a crossover to crystalline wires with shell-closings corresponding to the completion of additional atomic layers. This observation completes the analogy between shell effects observed for clusters and nanowires.}
\end{abstract}

\vskip2pc]

\narrowtext

\newcommand{\av}[1]{\mbox{$\langle #1 \rangle$}}

The remarkable achievements in microelectronics over the past decades go hand in hand with our ability to miniaturize electronic circuits further and further. Yet it is well known that there will come a moment when integrated circuitry will become so small it will no longer function by the laws of conventional electronics. Ultimately, a metallic system whose size is comparable to that of a single atom becomes essentially quantum mechanical, leading to a dramatic change in its properties. And we are getting close to this limit: the newest technology permits building electronic components with lateral dimensions of only a few hundreds of atoms. Anticipating even further miniaturization, it becomes very important to understand the electronic and mechanical properties of quantum conductors of atomic dimensions. 

Perhaps the simplest system for studying these novel properties is a point contact of an alkali metal. Alkali metals such as sodium or potassium can be viewed as almost ideal free-electron systems, where simple theoretical models are well applicable for the quantitative description of many fundamental phenomena. Therefore, although not nearly as suitable for chip fabrication as aluminum or copper, these metals constitute ideal test systems for gaining insight into the behavior of atomic-scale conductors. The use of scanning tunneling microscopes (STM) \cite{agrait93,pascual93,olesen94} and related techniques, such as mechanically controllable break junctions (MCBJ) \cite{muller92}, facilitates the study of atomic-size metallic conductors. The MCBJ technique, schematically depicted in the inset in Fig.\,1, is particularly useful for studying the behavior of nanowires made of the extremely reactive alkali metals. A bulk piece of such metal is mounted in a vacuum can, cooled down to liquid helium temperatures and then broken in order to expose clean fracture surfaces. These clean surfaces can be brought in and out of contact using a piezoelectric element for fine positioning control. By pulling out a contact into a thin bridge, or a nanowire, and simultaneously measuring the conductance, we can monitor its evolution all the way down to the point when it consists of just a few atoms and finally breaks completely 
(Fig.\,1). By pressing the electrodes back together a new contact can be made and the whole process can be repeated over and over again.

\begin{figure}[!b]
\begin{center}
\leavevmode
\epsfig{figure=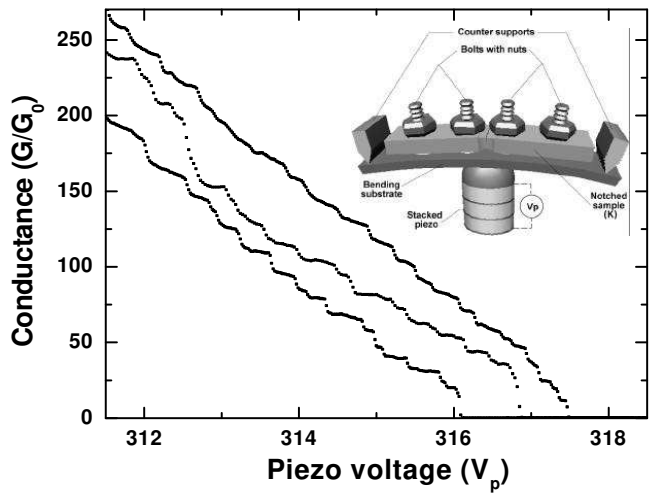,width=8.5cm}
\end{center}
\caption{Typical conductance traces recorded during the breaking of potassium contacts at a sample temperature of 100 Kelvin. The piezo voltage is linearly proportional to the distance between the electrodes. The conductance was measured using a standard 4-point dc technique with 1 mV voltage bias. A 16 bit analog-to-digital converter reads the signal at the output of a current-to-voltage converter, and the data acquisition and analysis is further handled by a pc. Each trace takes 0.1 seconds. Inset: sample mounting for an alkali break-junction. This setup is mounted inside a vacuum can immersed in liquid helium. The cryogenic vacuum inside the can prevents contact contamination even when the sample is locally heated to 100 K. The total length of the substrate is 22 mm.}
\label{fig:traces}
\end{figure}

Initially comprised of many atoms, a metallic contact can be viewed as a combination of the atomic structure, i.e. the exact configuration of atoms in the contact, and the electronic system, which defines the energy spectrum and electronic transport. Using the semi-classical expression for the conductance $G$ of a ballistic nanowire \cite{torres94,hoppler98},
\begin{equation}
G=gG_0\simeq G_0 \bigl[ \bigl( {k_{\rm F}R\over 2} \bigr)^2 - {k_{\rm F}R\over 2} + {1 \over 6} + \cdots \bigr] ,
\label{eq:g-expansion}
\end{equation}
where $G_0=2e^2/h$ is the quantum unit of conductance and $k_{\rm F}$ is the Fermi wave vector, we can relate the reduced conductance $g$ to the radius $R$ of the narrowest cross-section of the contact. Thus, by measuring the conductance during elongation we can observe the thinning of the narrowest part of the contact. Typical conductance traces, recorded as a function of contact elongation, consist of many slightly inclined plateaus with abrupt vertical jumps in between (Fig.\,1). A conductance plateau indicates the build-up of the elastic strain in the atomic structure, which is subsequently released during an atomic rearrangement, leading to a jump in the conductance and a discontinuous decrease in the radius of the contact. During each rearrangement many atoms change their positions to create another ``frozen'' meta-stable atomic configuration. Since the number of configurational degrees of freedom in a contact is very large and we have no direct control over the rearrangement process, it is highly improbable that the same conductance trace is repeated in different elongation cycles. Therefore, although each conductance trace is still shaped like a staircase, they all are rather irregular and vary in details.

However, we can look for common features by statistically analyzing conductance traces from many elongation cycles and compiling the conductance values in a histogram. A peak on such conductance histogram would then indicate the ``preferred'' conductance value. This may be the effect of conductance quantization \cite{krans95}, or, for larger contacts, correspond to a stable contact configuration when the system finds itself in a local energy minimum. Thus a histogram effectively filters out all irreproducible meta-stable atomic configurations and reveals the diameters at which a nanowire is exceptionally stable. 

In previous experiments we have shown that increasing the temperature to a sizeable fraction of the melting point reduces further the undesirable effect of ``frozen'' configurations and restores the almost perfect jellium-like behavior of sodium. The ionic structure adjusts itself to the configurations of minimal energy determined solely by the electronic system. For such contacts one observes a large number of distinct peaks in the histogram, periodic when plotted as a function of the square root of the conductance (which is almost linearly proportional to the radius of the contact (Eq.\,1)). It was shown in \cite{yyr1,yyr2} that this periodic pattern could be explained in terms of an electronic shell effect, which produces periodic minima in the thermodynamic potential of the nanowires as a function of their radius. Diameters with minimal free energy will be more frequently encountered during the pulling of the wires, giving rise to the maxima in the histograms. The same mechanism determines the stability of clusters of metal atoms produced in vapor jets in vacuum \cite{deheer93,knight84}, where the numbers of atoms required to complete an electronic shell are called "magic numbers". In the present investigation of magic wire radii for alkali metal nanowires we have encountered a new periodic structure at larger diameters, which points at a transition from electronic shell effect to {\it atomic shell structure}, in close analogy to what has been observed for the clusters \cite{martin96}.

\begin{figure}[!t]
\begin{center}
\leavevmode
\epsfig{figure=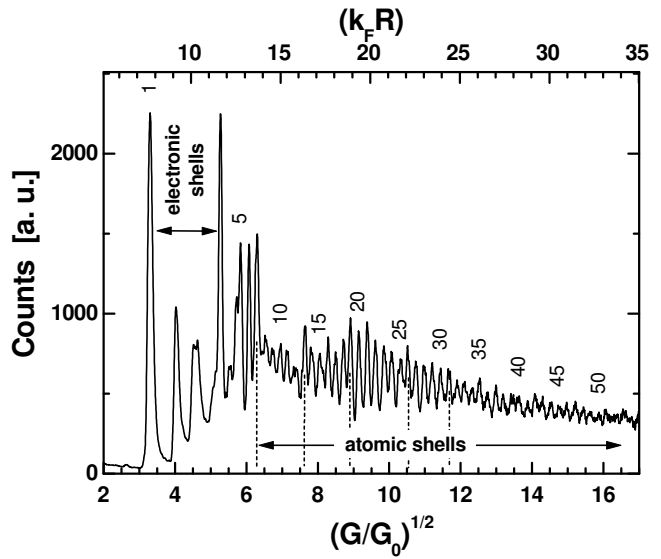,width=8.5cm}
\end{center}
\caption{ Histogram of conductance values obtained from 1100 individual conductance curves recorded while stretching the contact between two pieces of potassium, using the mechanically controllable break junction technique. In order to bring out the periodicity of the structure the histogram is plotted as a function of the square root of the conductance, while the latter is given in units of the conductance quantum, $G_0 =2e^2/h=(12.9\,{\rm k}\Omega)^{-1}$.  The temperature of the sample is 100 K.}
\label{fig:histogram}
\end{figure}

A conductance histogram for potassium, recorded at about 1/3 of its melting temperature, is shown in Fig.\,2. Two overlapping sets of periodic oscillations can be identified: one is found up to a reduced conductance $g$ of about 36 ($\sqrt{g} \le 6$) and the other up to 260 ($\sqrt{g} \simeq 16$). The first one has the period of the electronic shell effect reported in \cite{yyr1}, but the second is a new oscillating phenomenon having approximately a three times smaller period. The position of the transition region between the two periodic features depends strongly on experimental parameters such as the sample temperature, the voltage bias and the depth of indentation. 
In Fig.\,3 the positions of the peaks of the histogram (Fig.\,2) are plotted against their serial index. The indices of the peaks are also shown in the histogram of Fig.\,2, where they are labeled in increments of five. The first few points in Fig.\,3, corresponding to the pronounced peaks in Fig.\,2, are fitted by a straight line with an approximate slope of $0.62\pm0.05$, which agrees with the slope of $0.56\pm0.01$ determined for the electronic shell effect in sodium nanowires \cite{yyr1}. By varying the experimental conditions one can greatly extend the conductance range over which this electronic shell structure is observed, while the slope always agrees with the periodicity expected for this effect. In the case of Fig.\,3, beyond these first few points up to 48 points obey a linear relation with a much smaller slope of $0.223\pm 0.001$. This new periodic structure has been reproduced for about 10 different contacts for several samples of the alkali metals K and Na. For K, the transition between these two sets of oscillations (electronic shell structure and the new pattern) is quite sharp, although the higher frequency oscillation extends a bit with reduced amplitude into the lower conductance range of the electronic shell region. For sodium the crossover is found at larger radii than for potassium, and for lithium only electronic shell structure is observed.

When searching for an explanation for this anomalous periodic structure, we are led by the research on metal clusters. For alkali metal clusters produced in vacuum a clear transition has been observed between a series of magic numbers given by the filling of electronic shells, and a different series of magic numbers determined by the closing of geometric shells of atoms \cite{deheer93,martin96,brack93}. The former effect, resulting from the quantization of energy levels in symmetrically confined systems, becomes weaker and eventually disappears as the size of the clusters increases. The latter series results from the fact that larger clusters have a highly symmetric crystalline shape and their surface energy attains a minimum when a complete layer of atoms covers the surface. We are therefore led to the assumption that larger nanowires have a similar tendency to order and assume crystalline axially symmetric shapes. Indeed, such stable hexagonal prisms with six close-packed facets were recently seen in a transmission electron microscope during the thinning of gold nanobridges by irradiation with a high-intensity electron beam \cite{kondo97}. 

Since the periodic pattern in Fig.\,2 extends to large wire diameters, we start from the assumption that the lattice structure in the wire is that of the bulk metal. The bulk lattice structure of potassium is body centered cubic (bcc) and the lowest energy surfaces are the [110] surfaces. A wire with exclusively [110] facets can be formed with its axis along [100] or [111]. The cross section for the former would be square while that of the latter is a hexagon. The hexagonal wire clearly has a smaller surface area and therefore would be more energetically favorable. The proposed structure is illustrated in the inset in Fig.\,3. 

\begin{figure}[!b]
\begin{center}
\leavevmode
\epsfig{figure=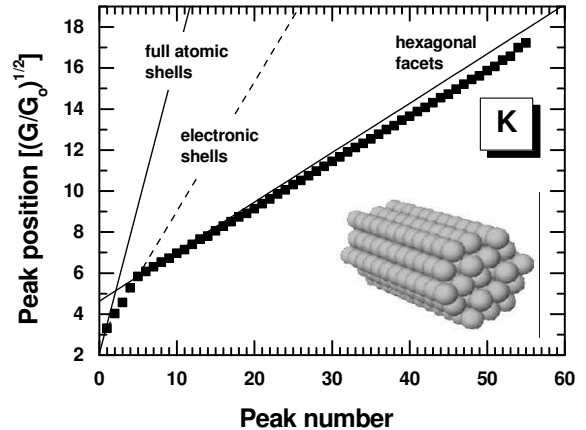,width=7.5cm}
\end{center}
\caption{The positions of the peaks in Fig.\,2 plotted against their sequentially numbered index. The upper line gives the slope expected for full atomic shell coverage of a nanowire with bcc arrangement of the atoms and the axis along [111]. The lower solid line has a slope of one-sixth of that of the former, and corresponds to the successive filling of individual facets of a hexagonal wire. A dashed line with a slope corresponding to {\it electronic} shell structure connects the lowest three data points. Inset: Illustration of the proposed structure of a hexagonal nanowire with the axis along the [111]-direction and the bulk bcc stacking.}
\label{fig:line}
\end{figure}

Let us now consider what would be the slope of $\sqrt{g}$ versus the peak index $m$, if we assume that the closing of the shells of atoms around such nanowire leads to peaks on the conductance histogram. The relation between the conductance and the wire cross-section $A$ can be approximated by the lowest order term, $g\simeq {k_{\rm F}^2 A \over 4 \pi}$, which is valid for $ k_{\rm F}R \gg 1$. The distance between [110] atomic layers is $d=a_0/\sqrt{2}$, where $a_0$ is the size of the cubic unit cell, and thus the slope becomes
\begin{equation}
\alpha = {{\rm d}\sqrt{g} \over {\rm d}m} = {3^{1/4}\over 2\sqrt{\pi}} k_{\rm F}a_0 .
\label{eq:slope}
\end{equation}
For a free-electron metal with bcc lattice $ k_{\rm F}a_0=(6\pi^2)^{1/3}$ and so we obtain the value $\alpha = 1.447$, shown in Fig.\,3 by a straight line labeled ``full atomic shells''. Clearly, this is a much higher slope than the one we observe, even higher than the one for the electronic shell effect. However, when we assume that a stable configuration is obtained each time a single {\it facet} of the hexagonal prism is completely covered with atoms, in analogy to what is observed for some metal clusters (e.g. for aluminum \cite{martin96,martin92}), then the slope becomes a factor 6 lower, equal to 0.241, in close agreement with the experimental data. The small deviation from the experimental data points may be related to a reduction of the perfect conductance of the nanowires by scattering on defects.

Support for this interpretation comes from considering the conductance of a wire with $m$ completely full shells of atoms. We obtain $\sqrt{g}=\alpha(m+m_0)$, where $m_0\simeq 0.5$ is an offset value, which depends somewhat on the boundary conditions for the electrons. From this expression, and using the experimental value $\alpha = 6\times 0.223=1.34$, we expect the closing of full shells at $\sqrt{g}=6.02$, 7.36 and 8.70, for $m$\,=\,4, 5 and 6, respectively. Apart from a small shift, these values correspond to the higher-intensity peaks with indices 7, 13 and 19, marked by dashed lines in Fig.\,2. The small shift may be absorbed in a slightly modified offset value $m_0$\,=\,0.66 for a more realistic confining potential for the electrons. These higher-intensity peaks are separated by 5 lower-intensity ones, which correspond to single filled facets in this interpretation. However, this periodic difference in peak amplitudes is not always clearly visible. Further evidence for 6-fold atomic packing comes from the distances between the peaks in the histogram. Each time another shell has been completed, the contribution of each facet to the cross-section of the nanowire should increase by roughly one atom, raising the conductance by approximately 1\,$G_0$ per complete facet. By examining the distance between the peaks (averaged over two neighboring peak distances) in Fig.\,2 we see that it indeed tends to increase by 1\,$G_0$ after each 6 peaks, as expected for the filling of the next hexagonal atomic layer.

Thus, we have evidence for two sets of ``magic numbers'': electronic and atomic. These two sets of oscillations in the conductance histogram compete with each other, just like it is the case in cluster physics \cite{brack93}. One shell-closing effect is related to the energy of the total volume of electrons, for which the amplitude of the oscillations in the thermodynamic potential decreases as $1/R$.  The other is due to the surface energy, for which the amplitude of the oscillations is roughly constant as a function of $R$. The transition between them depends on the parameters of the experiment. The atomic shell oscillations are observed at larger diameters (conductances) than the electronic ones, but they may overlap substantially. For Li, and in many cases for Na (see \cite{yyr1,yyr2}), the electronic shell structure oscillations completely dominate the spectrum. 

The periodic peak structure in the histograms is only observed at temperatures well above helium temperature. The thermal energy is required in order to have sufficient mobility of the atoms allowing the structure to accommodate to the lowest free energy. Potassium has the lowest melting temperature among the three alkali metals (Li, Na, K) studied by us. This means that at a given temperature its atoms have the highest mobility and for this metal we obtain the largest number of oscillations in the conductance histogram. 

From our data we cannot exclude other atomic wire arrangements. For example, Kondo and Takayanagi \cite{kondo00} have observed spectacular helical arrangements in nanowires of gold, and similar unusual atomic arrangements were found in computer simulations of Au, Al and Pb atomic wires \cite{gulseren98,wang01,tosatti01}. However, in all cases beyond a critical radius of the order of three atomic distances, the bulk lattice structure is recovered. Therefore it is likely that the nanowires with the diameters in the range of the atomic shell structure oscillations observed here have a regular atomic stacking structure. On the other hand, the alkali metals Li and Na have low temperature martensitic phase transitions toward a close-packed atomic structure and it is possible that the surface tension increases further the tendency towards close packing in the nanowires. It turns out that one can construct a close-packed nanowire of similar shape to the bcc structure proposed above. Such wires would have a face centered cubic lattice, with the axis along [011] and six facets perpendicular to $[100], [1{\bar 1}1], [{\bar 1}{\bar 1}1], [{\bar 1}00], [{\bar 1}1{\bar 1}],$ and $[11{\bar 1}]$. For this arrangement we obtain $\alpha=3^{5/6}\pi^{1/6}/2^{13/12}=1.427$, which is very close to the value for the bcc structure. With present experimental accuracy it is not possible to distinguish between these structures in our nanowires. We conclude that the data present evidence for spontaneous formation of atomically ordered wires for alkali metals K and Na at about 100 K, which have six facets of approximately equal width.

\end{document}